\documentclass[
aps,%
12pt,%
final,%
notitlepage,%
oneside,%
onecolumn,%
nobibnotes,%
nofootinbib,%
superscriptaddress,%
noshowpacs,%
centertags]%
{revtex4}
\usepackage{lineno,hyperref}
\modulolinenumbers[5]

\usepackage{amsmath}
\usepackage[pdftex]{graphicx}
\renewcommand{\Re}{\mathop{\mathrm{Re}}}
\renewcommand{\Im}{\mathop{\mathrm{Im}}}
\begin{document}

\title{On the spreading width of the Isobaric Analog Resonances}

\author{\firstname{G.~V.}~\surname{Kolomiytsev}}
\affiliation{National Research Nuclear University "MEPhI", Moscow 115409, Russia}
\author{M.~L.~Gorelik}
\affiliation{Moscow Economic School, Moscow 123022, Russia}
\author{M.~H.~Urin}
\affiliation{National Research Nuclear University "MEPhI", Moscow 115409, Russia}

\begin{abstract}
A description of the spreading width of the Isobaric Analog Resonances in medium-heavy mass spherical nuclei is proposed within a combined approach. This latter consists in incorporating the ``Coulomb description'' of isospin-forbidden processes into the newly developed particle-hole dispersive optical model. Within this approach, the observed spreading width of the resonances based on the $^{208,209}$Pb parent-nuclei ground state is quantitatively estimated without the use of specific adjustable parameters.
\end{abstract}

\maketitle

\section{Introduction}
A small value (several tens keV) of the spreading width of the Isobaric Analog Resonances (IARs), $\Gamma^{\downarrow}_{A}$, is the impressive manifestation of the approximate isospin-symmetry conservation in medium-heavy mass nuclei. The spreading width of an arbitrary giant resonance (including the IAR) is determined by coupling of corresponding particle-hole-type excitations to many-quasiparticle configurations (chaotic states). In the case of the IAR, this coupling is significantly suppressed and realized only due to isospin mixing. In medium-heavy mass nuclei, the main mixing mechanism consists in IAR coupling to its overtone (the Isovector Monopole Giant Resonance in the $\beta^{-}$-channel (IVMGR$^{(-)}$)) via a variable part of the mean Coulomb field (see, e.g., Ref. \cite{Auerbach}).

A realistic attempt to estimate quantitatively the IAR spreading width has been undertaken rather recently \cite{GRU_2010} within the approach, that includes the "Coulomb description" of IAR properties and consideration of the spreading effect on properties of giant resonances, having the "normal" isospin, within a semi-microscopic model \cite{Urin_NPA_2008}. The shortcoming of this model is a non-correct description of mentioned giant resonances at their distant "tails" (the IAR is located at the low-energy "tail" of the IVGMR$^{(-)}$). In the present work, for the description of the spreading effect we apply the newly developed particle-hole dispersive optical model \cite{Urin_PRC_2013}, which is free from the above-mentioned shortcoming. In Sect. 2, we present the basic relationships used for the description of the IAR spreading width within the proposed approach. The choice of model parameters, the calculation results obtained for the IARs based on the $^{208, 209}$Pb parent-nuclei ground state, and comparison with the corresponding experimental data are given in Sect. 3. Conclusive remarks and perspectives for further studies of IAR damping by the use of the proposed approach are contained in Sect. 4.
\section{"Coulomb description" of IAR damping}
\subsection{Coulomb strength function and the IAR total width}
Existence and properties of the IARs are closely related to the approximate isospin-symmetry conservation in nuclei. Let $\hat{H}$ be a model Hamiltonian that includes the mean Coulomb field $\widehat{U}_{C} = \frac{1}{2} \sum \limits_{a}(1 - \tau^{(3)}_a) U_C(r_{a})$. In medium-heavy mass nuclei, a variable (in space) part of this field is mainly responsible for isospin-symmetry violation (see, e. g., Ref. \cite{Auerbach}). In such a case, the equation of motion for the Fermi operator $\widehat{T}^{(-)} = \sum \limits_{a} \tau^{(-)}_{a}$, that generates proton-neutron-hole $(p\bar{n})$ monopole excitations associated with the IAR, can be presented in the form \cite{GRU_2010}:
\begin{equation}\label{eq_commut}
[\widehat{H}, \widehat{T}^{(-)}] -\Delta_{C}\widehat{T}_{(-)} = \widehat{V}^{(-)}_{C}, \quad \widehat{V}^{(-)}_{C} = \sum \limits_{a} \left( U_{C}(r_a)-\Delta_C \right) \tau^{(-)}_a,
\end{equation}
where the parameter $\Delta_{C}$ is defined below. This equation allows one to get a correspondence between the Fermi and Coulomb energy-averaged strength functions $S^{(-)}_{F}((\omega)$ and $S^{(-)}_{C}(\omega)$, respectively:
\begin{equation}\label{eq_SF-SC}
S^{(-)}_{F}((\omega) = \frac{S^{(-)}_{C}(\omega)}{|\omega - \Delta_C|^2}.
\end{equation}
Here, $\omega$ is the excitation energy counted off the energy of the parent-nucleus ground state, the Fermi and Coulomb energy-averaged strength functions are related to the monopole probing operators (external fields) $V^{(-)}_{F}(x) = \tau^{(-)}$ and $V^{(-)}_{C}(x) = (U_{C}(r) - \Delta_{C})\tau^{(-)}$, respectively.

In the experimental excitation functions of $(pp')$- and $(pn_{tot})$-reactions, the IAR is found as a narrow well-formed resonance. For this reason, the Fermi strength function in a vicinity of IAR can be parameterized by a Lorenzian:
\begin{equation}\label{eq_SF_lorentz}
S^{(-)}_{F} = \frac{1}{2\pi}\frac{S_A \Gamma_A}{|\omega - \omega_A + i\Gamma_A /2|^2},
\end{equation}
where $S_{A}$ is the IAR Fermi strength (close to the neutron excess), $\omega_{A}$ and $\Gamma_{A}$ are, respectively, the IAR excitation energy and total width. As follows from Eqs. (\ref{eq_commut}) --- (\ref{eq_SF_lorentz}), the IAR total width is determined by the, generally speaking, transcendental equation:
\begin{equation}\label{eq_width_SC}
\Gamma_{A} = \frac{2 \pi}{S_A} S^{(-)}_{C}(\omega_A).
\end{equation}
In this equation, the Coulomb strength function is defined with the use of the complex-valued quantity $\Delta_{C} = \omega_{A} - (i/2)\Gamma_{A}$. As a function of $\omega$, the mentioned strength function exhibits the maximum, corresponding to the IVGMR$^{(-)}$. Considering this resonance as a Lorenzian, one gets from Eq.(\ref{eq_width_SC}) the known qualitative estimation of the IAR total width: $\Gamma_{A} = \beta^2_{M, A} \Gamma_{M}(\omega_A)$ \cite{Auerbach}. Here, $\beta_{M, A}$ is the amplitude of IAR and IVGMR$^{(-)}$ isospin mixing caused by a variable part of the mean Coulomb field, $\Gamma_{M}$ is the IVGMR$^{(-)}$ total width taken at the IAR energy. Such an estimation shows that the quantitative description of the IAR total width and its main components, the proton-escape width $\Gamma^{\uparrow}_{A}$ and spreading width $\Gamma^{\downarrow}_{A}$  ($\Gamma^{\uparrow}_{A} + \Gamma^{\downarrow}_{A} = \Gamma_{A}$), needs the correct description of the distant low-energy "tail" of the IVGMR$^{(-)}$. Hereafter, we neglect by IAR "rare" decays (such as radiation and direct-neutron decays).

As any strength function taken at the energy, that exceeds the nucleon separation energy, the Coulomb strength function can be divided into the direct (one-nucleon escape) and spreading (statistical) parts:
\begin{equation}\label{eq_SC_stat}
S^{(-)}_{C}(\omega) = S^{(-), \uparrow}_{C}(\omega) + S^{(-), \downarrow}_{C}(\omega).
\end{equation}
These parts describe, respectively, direct proton and statistical (mainly neutron) decays of high-energy monopole $(p\bar{n})$-type states. As follows from Eqs.(\ref{eq_width_SC}),(\ref{eq_SC_stat}), the main components of the IAR total width, $\Gamma^{\uparrow}_{A}$ and $\Gamma^{\downarrow}_{A}$, are determined by the respective components of the Coulomb strength function taken at the IAR energy.

\subsection{Coulomb strength function within the PHDOM}
Abilities to describe direct-decay properties and distant "tails" of various giant resonances are related to the specific features of the newly developed  particle-hole dispersive optical model (PHDOM) \cite{Urin_PRC_2013}. Being an extension of the standard \cite{Urin_PRC_2013} and nonstandard \cite{Urin_NPA_2008} continuum-RPA (cRPA) versions to a phenomenological (and in average over the energy) description of the spreading effect in closed-shell nuclei, the model includes a few ingredients. They are: (i) the Landau-Migdal p-h interaction $F(x_{1}, x_{2}) \rightarrow 2F' \vec{\tau}_{1} \vec{\tau}_{2} \delta(\mathbf{r}_{1} - \mathbf{r}_{2})$ (below  charge-exchange monopole (p-h)-type excitations are only considered); (ii) a realistic phenomenological mean field, in which essential for the description of  above-mentioned excitations the symmetry potential and mean Coulomb field are evaluated self-consistently (see, e.g., Ref. \cite{KIU_2014}); (iii) the imaginary and real (dispersive) parts, respectively $W(\omega)$ and $P(\omega)$, of the strength of the phenomenological energy-averaged specific p-h interaction (p-h self-energy term) responsible for the spreading effect. Below the PHDOM basic equations are first given in applying to the description of high-energy charge-exchange monopole excitations, having "normal" isospin.

Let $V^{(-)}(x) = V(r)\tau^{(-)}$ be a monopole Fermi-type probing operator (external field), in which the radial part $V(r)$ might be complex-valued quantity. The energy-averaged strength function $S^{(-)}_{V}(\omega)$ and polarizability $P^{(-)}_{V}(\omega)$, corresponding to this operator, are determined by an effective field $ \widetilde{V}(r, \omega)$. After separation of spin-angular and isospin variables, one gets the expressions for the above-mentioned quantities and the equation for the effective field:
\begin{equation}\label{eq_SV_ImP}
S^{(-)}_{V}(\omega) = -\frac{1}{\pi} \Im P^{(-)}_{V}(\omega),
\end{equation}
\begin{equation}\label{eq_P}
P^{(-)}_{V}(\omega) = \int V^{*}(r) A^{(-)}(r,r',\omega)  \widetilde{V}(r', \omega) drdr',
\end{equation}
and
\begin{equation}\label{eq_V_eff}
\widetilde{V}(r, \omega) = V(r) + \frac{F'}{2\pi r^2} \int A^{(-)}(r,r',\omega)  \widetilde{V}(r', \omega) dr'.
\end{equation}
Here, $A^{(-)}(r, r', \omega)$ is the radial monopole component of the "free" p-h propagator in the $\beta^{(-)}$-channel. The expression for this component, obtained within the PHDOM with taking approximately the single-particle continuum into account \cite{Urin_PRC_2013}, can be presented as the sum:
\begin{eqnarray}\label{eq_A_sum_components}
	\begin{aligned}
A^{(-)} = A^{(-)}_{1} + A^{(-)}_{2} +  A^{(-)}_{3},\\
A^{(-)}_{1}(r,r',\omega) = \sum \limits_{\nu,(\pi)} t^2_{(\pi)(\nu)} n_{\nu} \chi_{\nu}(r)\chi_{\nu}(r')g_{(\pi)}(r,r',\epsilon_{\nu}+\omega),\\
A^{(-)}_{2}(r,r',\omega) = \sum \limits_{(\nu),\pi} t^2_{(\pi)(\nu)} n_{\pi} \chi_{\pi}(r)\chi_{\pi}(r')g_{(\nu)}(r,r',\epsilon_{\pi}-\omega),\\
A^{(-)}_{3}(r,r',\omega) = \sum \limits_{\nu,\pi} t^2_{(\pi)(\nu)} n_{\pi} n_{\nu} \chi_{\pi}(r)\chi_{\pi}(r')\chi_{\nu}(r)\chi_{\nu}(r') a_{\pi \nu}(\omega),\\
a_{\pi \nu} (\omega) = \frac{2\left( iW(\omega) - P(\omega) \right) f_{\pi}f_{\nu}}{(\epsilon_{\pi} - \epsilon_{\nu} -\omega)^2-\left( iW(\omega) - P(\omega) \right)^2 f^2_{\pi}f^2_{\nu}}.
	\end{aligned}
\end{eqnarray}
Here, the bound-state radial wave functions $\chi_{\mu}(r)$ satisfy the equation $(h_{(\mu)}(r) - \epsilon_{\mu})\chi_{\mu}(r) = 0$, where $\mu$ is the set of single-particle quantum numbers $n_{r}, j, l$ ($(\mu) = j, l$) for neutrons ($\mu = \nu$) and protons ($\mu = \pi$), $h_{(\mu)}(r)$ is the radial part of a single-particle Hamiltonian (this part includes the spin-orbit and centrifugial terms); $n_{\mu} = N_{\mu}/(2j_{\mu}+1)$ is the occupation number ($N_{\mu}$ is the number of nucleons filling the single-particle level $\mu$); $t^{2}_{(\pi)(\nu)} = (2j_{\nu}+1)\delta_{(\pi)(\nu)}$ is the squared kinematical factor; the optical-model-like Green functions $g_{(\pi)}(r, r', \epsilon_{\nu} +\omega)$ and $g_{(\nu)}(r,r', \epsilon_{\pi} -\omega)$ satisfy the equations:
\begin{eqnarray}\label{eq_sp_gf_prot}
	\begin{aligned}
\left\{ h_{(\pi)}(r) - \left[ \epsilon_{\nu} + \omega + \left( iW(\omega)-P(\omega) \right) f_{\nu} f(r) \right]\right\} g_{(\pi)}(r,r',\epsilon_{\nu}+\omega) = -\delta(r-r'),\\
\left\{ h_{(\pi)}(r) - \left[ \epsilon_{\nu} + \omega + \left( iW(\omega)-P(\omega) \right) f_{\nu} f(r) \right]\right\} \chi_{\epsilon,(\pi)}(r) = 0;\\
	\end{aligned}
\end{eqnarray}
\begin{equation}\label{eq_sp_gf_neut}
\left\{ h_{(\nu)}(r) - \left[ \epsilon_{\pi} - \omega + \left( iW(\omega)-P(\omega) \right) f_{\pi} f(r) \right]\right\} g_{(\nu)}(r,r',\epsilon_{\pi}-\omega) = -\delta(r-r').
\end{equation}
We show here also the equation for the proton optical-model-like continuum-state wave functions $\chi_{\epsilon,(\pi)}(r)$ ($\epsilon = \epsilon_{\nu} + \omega > 0$), having standing-wave asymptotical behavior. These wave functions (in the limit $W=P=0$, these are normalized to $\delta$-function of the energy) enter in the definition of the above-mentioned proton-escape Coulomb strength function
\begin{eqnarray}\label{eq_SC_sp_wf}
	\begin{aligned}
S^{(-), \uparrow}_{C}(\omega) =& \sum_{\nu} S^{(-), \uparrow}_{C,\nu}(\omega),\\
S^{(-), \uparrow}_{C,\nu}(\omega) =& N_{\nu} \delta_{(\pi)(\nu)}
\left|	\int \chi^{*}_{\epsilon, (\pi)}(r) \widetilde{V}^{(-)}_C(r,\omega) \chi_{\nu}(r) dr \right|^2
	\end{aligned}
\end{eqnarray}
that determines the IAR total proton-escape width $\Gamma^{\uparrow}_{A} =  \frac{2\pi}{S_{A}} S^{(-), \uparrow}_{C}(\omega_{A})$. Finally, we get the expression for the IAR spreading width
\begin{equation}\label{eq_spread_width}
\Gamma^{\downarrow}_{A} = \Gamma_{A} - \Gamma_{A}^{\uparrow}
\end{equation}
in terms of the proper Coulomb strength functions, as it follows from Eqs. (\ref{eq_width_SC}), (\ref{eq_SC_stat}), (\ref{eq_SC_sp_wf}).

In ignoring the spreading effect ($W(\omega) = P(\omega) =0$ in Eqs. (\ref{eq_A_sum_components})---(\ref{eq_sp_gf_neut})), when $\Gamma^{\downarrow}_{A} = 0$, the IAR energy $\omega_{A, 0}$ and Coulomb polarizability $P^{(-)}_{C, 0}(\omega_{A, 0})$ can be evaluated within the cRPA. That allows one to get an estimation of the (relatively small) IAR spreading shift by the relationship:
\begin{equation}\label{eq_omega_shift}
\omega_{A} - \omega_{A, 0} = \frac{1}{S_A} \Re \left\{ P^{(-)}_{C}(\omega_A) - P^{(-)}_{C,0}(\omega_{A, 0}) \right\}.
\end{equation}
This statement completes presentation of the approach to a quantitative description of the IAR spreading width for medium-heavy mass closed-shell and closed-shell+valence-neutron parent nuclei.

In conclusion of this Section, we note, that in applying to the description of high-energy $(n\bar{p})$-type monopole excitations, the PHDOM basic equations can be obtained from Eqs. (\ref{eq_SV_ImP}) --- (\ref{eq_sp_gf_neut}) by the substitution $\pi \leftrightarrow \nu$, or (that is the same) $\omega \rightarrow -\omega$. In the so obtained equations, $\omega$ means the excitation energy counted off the parent-nucleus ground-state energy. The above-mentioned equations can be used, in particular, for description of the IVGMR$^{(\mp)}$ strength functions, $S^{(\mp)}_M(\omega)$.

\section{Choice of model parameters. Calculation results}
\begin{figure}[ht]
  \includegraphics[width=1.0\linewidth]{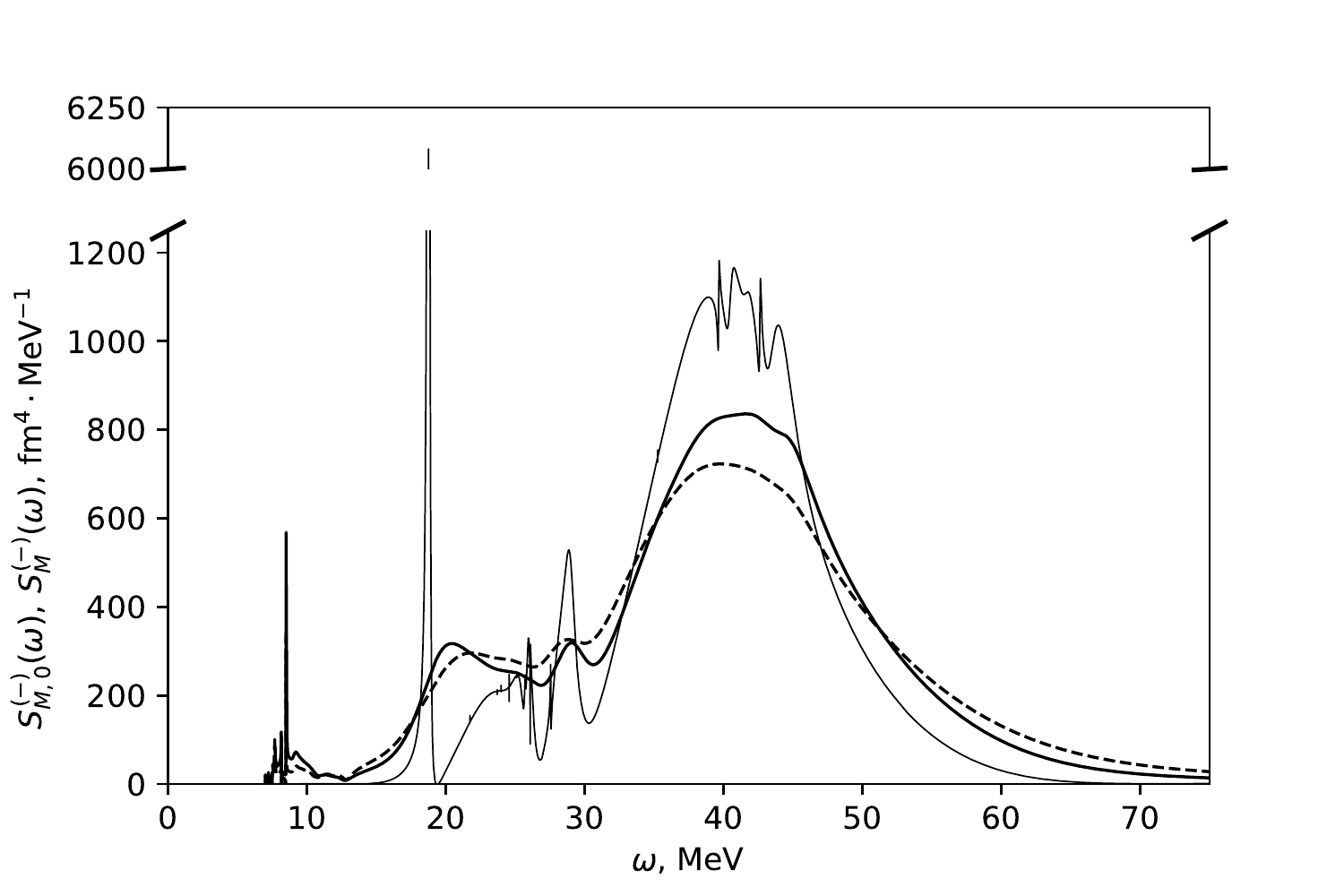}
  \caption{\label{pict_SMonopole} The monopole strength functions $S^{(-)}_M(\omega)$ and $S^{(-)}_{M,0}(\omega)$ (in fm$^4 \cdot$MeV$^{-1}$) calculated for the $^{208}$Pb parent nucleus within the cRPA (thin line) and PHDOM with the use of the ``traditional'' and new sets of the ``spreading'' parameters (dashed and bold lines, respectively).}
\end{figure}

As an example of implementations of the above-described approach, we consider below the spreading width of the IARs based on the ground state of the $^{208, 209}$Pb parent nuclei. In such a consideration, we turn first to ingredients of the PHDOM, which is the basic element of the proposed approach (Subsection II.B). A realistic partially self-consistent phenomenological mean field is described in details in Ref. \cite{KIU_2014}, where the list of mean-field parameters (including the Landau-Migdal parameter $f' = F'/(300 \text{ MeV} \cdot \text{fm}^{3}))$ for the $^{208}$Pb parent nucleus is also given. The parameterization of the phenomenological quantity, the imaginary part $W(\omega)$ (and, therefore, the expression for the dispersive real part $P(\omega))$ of the strength of the energy-averaged p-h self-energy term, is given in Refs. \cite{Urin_PRC_2013, TU_2009} for excitations in the neutral channel. In consideration of (p-h)-type excitations in the charge-exchange channels, we use the similarly parameterized quantity $W(E_{x})$, where $E_{x} = \omega - Q$ is the excitation energy, counted off the compound-nucleus ground-state energy ($Q$ is the difference of the ground-state energies of the corresponding compound and parent nuclei). Two sets of the ``spreading'' parameters (the strength $\alpha$ and ``saturation'' parameter $B$), which enter in the quantity $W(E_x)$, are chosen to describe within the PHDOM the monopole strength function $S^{(-)}_M (\omega)$ and, therefore, the IVGMR$^{(-)}$ energy and total width experimentally known for $^{208}$Pb parent nucleus with poor accuracy \cite{Errel_PRC_1986}. The first (``traditional'') set is close to that used previously for the description within the PHDOM of the low-energy giant resonances (isovector dipole \cite{tulupov2014description} and isoscalar monopole \cite{gorelik2016investigation}). The strength function $S^{(-)}_M (\omega)$ is calculated with the use of the probing operator $V^{(-)}_M(x)$ chosen to minimize the excitation of IAR: $V^{(-)}_M(x) = (r^{2} - \langle r^{2} \rangle)\tau^{(-)}$. Here, the brackets $\langle ... \rangle$ mean averaging over the neutron-excess density. The strength function $S^{(-)}_M (\omega)$ calculated within the PHDOM and cRPA for $^{208}$Pb are compared in Fig. \ref{pict_SMonopole}. As follows from this comparison, the single-particle continuum gives essential contribution to formation of the IVGMR$^{(-)}$. Two sets of ``spreading'' parameters $\alpha$ and $B$, the calculated and experimental energy and total width of the IVGMR$^{(-)}$ are given in Table \ref{table_gamma}. The use of the new set of adjustible parameters is found to be preferable.

After the above-described choice of model parameters, the IAR energy $\omega_{A}$ and total width $\Gamma_{A}$ can be evaluated within the approach, as follows. First, the corresponding cRPA equations are solved to calculate the Fermi strength function $S^{(-)}_{F, 0}(\omega)$ and then to evaluate the quantities $S_{A}$, $\omega_{A, 0}$ and $\Gamma_{A, 0}$. Using these quantities, one can calculate within the cRPA the Coulomb polarizibility $P^{(-)}_{C, 0}(\omega)$. Secondly, from the system of transcendental equations (\ref{eq_width_SC}) and (\ref{eq_omega_shift}) one can evaluate the IAR parameters $\omega_{A}$ and $\Gamma_{A}$ by means of an iterative procedure, which is well converged. These quantities are finally used to evaluate within the approach the IAR proton-escape and spreading widths in accordance with Eqs. (\ref{eq_SC_sp_wf}) and (\ref{eq_spread_width}), respectively. The calculated values $\Gamma^{\downarrow}_{A}$ given in Table \ref{table_gamma} for the IARs based on the ground state of the $^{208, 209}$Pb parent nuclei can be considered as a quantitative estimation of the corresponding experimental values \cite{Reiter_ZPA_1990}.

\begin{figure}[ht]
  \includegraphics[width=1.0\linewidth]{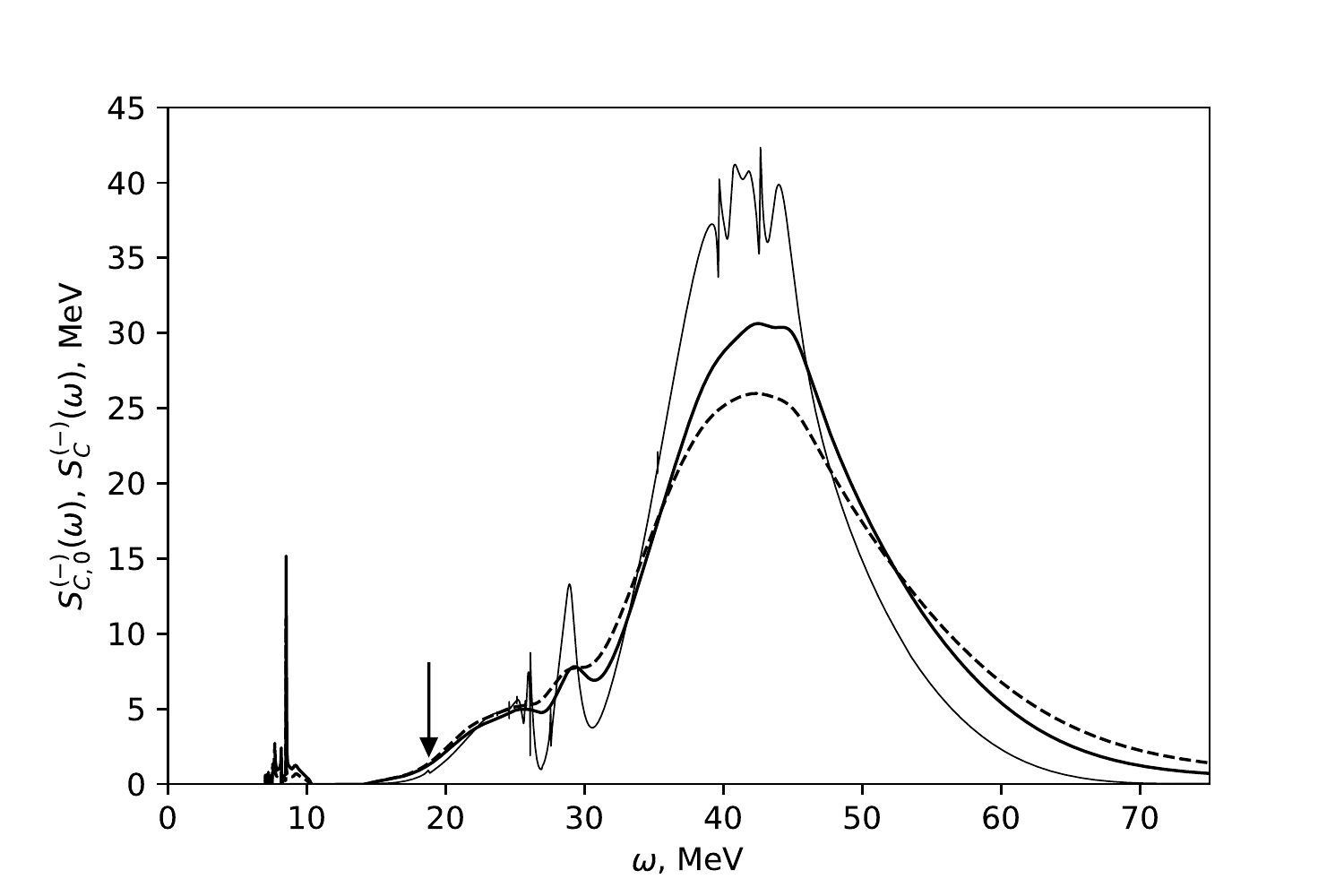}
  \caption{\label{pict_SCoul} Same as in Fig. \ref{pict_SMonopole} but for the Coulomb strength functions (in MeV). The arrow marks the IAR energy.}
\end{figure}

In conclusion of this Section, we show in Fig. \ref{pict_SCoul} the Coulomb strength functions $S^{(-)}_{C}(\omega)$ and $S^{(-)}_{C, 0}(\omega)$  calculated within the PHDOM and cRPA. Differently from the data shown in Fig. \ref{pict_SMonopole}, these strength functions exhibit no resonance structure in the IAR region. This point can be considered as an evidence of consistency of the proposed approach.
\section{Conclusive remarks. Perspectives}
As applied to the closed-shell and closed-shell + valence-neutron parent nuclei, we have used the newly developed particle-hole dispersive optical model for the description of high-energy charge-exchange monopole excitations. In particular, for the description of the Isobaric Analog Resonances we have proposed the approach, in which the ``Coulomb description'' of isospin-forbidden processes is incorporated into the above-mentioned model. As the first step in implementations of the approach, we have formulated the method for evaluation of the impressive quantity, the IAR spreading width. The method has been realized for the IARs based on the ground state of the $^{208, 209}$Pb parent nuclei, and a quantitative estimation of the corresponding experimental data has been obtained without the use of specific adjustable parameters.

The following steps in studying monopole charge-exchange excitations within the proposed approach might be the following. 1) A quantitative estimation of the IAR partial proton-escape widths with taking into account their sharp energy-dependence on the escaped-proton energy. 2) The description of the IAR asymmetry (determined by the so-called IAR mixing -phase) in the excitation functions of proton-induced reactions. 3) A quantitative estimation of the partial branching ratios for direct proton (neutron) decay of the Isovector Giant Monopole Resonance in the $\beta^{(-)}$ - ($\beta^{(+)}$ - ) channel. An extension of the approach in applying to medium-heavy mass spherical nuclei, having developed nucleon pairing, is also in sight.
\begin{acknowledgments}
This work is partially supported by the Russian Foundation for Basic Research under grant No. 15-02-080007 (G. V. K., M. L. G., M. H. U.), and by the Competitiveness Program of National Research Nuclear University ``MEPhI'' (G. V. K., M. H. U.).
\end{acknowledgments}
\bibliography{IAR-paper_KGU_biblio}
\begin{table}
\begin{tabular}{|c|c|c|c|c|c|}
\hline
  $\alpha$, MeV$^{-1}$ & $B$, MeV  & $\omega_M$, MeV & $\Gamma_M$, MeV & $\Gamma^{\downarrow}_{A}$, keV ($^{208}$Pb) & $\Gamma^{\downarrow}_{A}$, keV ($^{209}$Pb) \\
  \hline
  0.07 & 7 & 41.7 & 16.7 & 62 & 64 \\
  \hline
  0.035 & 14 & 39.8 & 19.4 & 73 & 75 \\
  \hline
  \multicolumn{2}{|c|}{Exp. data} & 37.0 $\pm$ 3.5  &  15.0 $\pm$ 6.0 & 78 $\pm$ 8 & 75 $\pm$ 7 \\
  \hline
\end{tabular}
\caption{\label{table_gamma} The ``traditional'' (first line) and new (second line) sets of the ``spreading'' parameters, and the calculated parameters of the IVGMR$^{(-)}$ in the $^{208}$Pb parent nucleus. The spreading width evaluated within the combined approach for the IARs based on the ground states of the $^{208}$Pb and $^{209}$Pb parent nuclei are given in last columns. The respective experimental data taken from Refs. \cite{Errel_PRC_1986} (IVGMR$^{(-)}$) and \cite{Reiter_ZPA_1990} (IARs) are also given.}
\end{table}
\end{document}